# Directional out-coupling from active hyperbolic metamaterials

*Tal Galfsky[1,2], Harish N.S. Krishnamoorthy[1,2], Ward D. Newman[3], Evgenii Narimanov[4], Zubin Jacob[3], Vinod M. Menon,[1,2*]*

[1]Department of Physics, Queens College, City University of New York (CUNY), Flushing, NY 11367, USA.

[2]Department of Physics, Graduate Center, CUNY, New York, NY 10016, USA.

[3]Department of Electrical and Computer Engineering, University of Alberta, Edmonton, AB T6G 2V4, Canada.

[4]Birck Nanotechnology Center, School of Computer and Electrical Engineering, Purdue University, West Lafayette, Indiana 47907, USA.

* Corresponding author: vmenon@qc.cuny.edu

**ABSTRACT** Hyperbolic Metamaterials (HMMs) have recently garnered much attention because they possess the ability for broadband manipulation of the photon density of states and sub-wavelength light confinement. However, a major difficulty arises with the coupling of light out of HMMs due to strong confinement of the electromagnetic field in states with high momentum called high-$k$ modes which become evanescent outside the structure. Here we report the first demonstration of directional out-coupling of light from high-k modes in an active HMM using a high index bulls-eye grating. Quantum dots (QDs) embedded underneath the metamaterial show highly directional emission through the propagation and out-coupling of resonance cones which are a unique feature of hyperbolic media. This demonstration of efficient out-coupling of light from active HMMs could pave the way for developing practical photonic devices using these systems.

Artificially engineered sub-structured materials having hyperbolic dispersion are known as hyperbolic metamaterials (HMMs) or indefinite media. Amongst the various metamaterial systems being explored, the HMMs have garnered most attention lately due to the simplicity of the structures, and a host of unique applications that span the frequency range from the microwave down to the UV while having non-magnetic permittivity ($\mu=1$). HMMs have been utilized in experimental demonstration such as nano-scale optical cavities [1], negative refraction in non-magnetic materials [2], super and hyper lenses [3,4], and control of spontaneous emission of quantum emitters [5–11]. All of these applications exploit sub-wavelength confinement of light in plasmonic Bloch modes in metal-dielectric structures [12]. However plasmonic modes become evanescent at the surface of the HMM and coupling of light into and out of these structures requires a mode matching scheme such as a high index substrate or nanopatterning [13,14]. Extracting light from the plasmonic modes of HMM is central to the development of active HMM devices such as sub-wavelength lasers and emitters. In this work we show for the first time experimental evidence for directional emission and light extraction from quantum dot embedded inside a HMM using a high-index bulls-eye grating.

A dipole in close proximity to a HMM emits light preferentially into the structure due to the enhanced photon density of states (PDOS) inside the HMM [5,9,15,16] with an emission pattern known as a resonance cone [12]. The resonance cone becomes evanescent at the interface between the HMM and air because of the large mode mismatch between the high-$k$ modes of the HMM and free space. However it is possible to translate the evanescent field into highly directional radiation in the far field using a grating coupler [17,18]. The ability to extract light from active HMM structures with directionality paves the way to practical applications of HMMs for realizing sub-wavelength lasers, single-photon sources, and sub-wavelength imaging.

Our HMM structure (shown schematically in Fig. 1a) consists of alternating layers of metal (Ag) and dielectric (Al2O3). In such a metal-dielectric composite, opposite signs of $\varepsilon_\parallel$ (x-y plane) and $\varepsilon_\perp$ (z-direction) can be achieved by controlling the fill fraction ( $ff$ ) of metal to dielectric. The fill fraction is defined as :

$$ff = \frac{t_m}{t_m + t_d} \quad (1)$$

where $t_m$ is the thickness of a metal layer and $t_d$ is the thickness of a dielectric layer. We have chosen silver (Ag) and alumina (Al$_2$O$_3$) as our building blocks with $ff = 0.5$ and thickness of each layer as 15 nm. The real and imaginary parts of the effective dielectric constants parallel and perpendicular to the layers as calculated using Effective Medium Theory (EMT) are shown in Fig. 1b. The opposite signs of the dielectric constants correspond to hyperbolic dispersion in the visible range of the spectrum. as The problem of mode mismatch with free space is illustrated schematically in the inset of Fig. 1b. The iso-frequency surface in air is a sphere with a smaller radius than the hyperbolic iso-frequency surface of the HMM and hence light cannot be coupled in and out of the structure from free space. The present work alleviates this issue by incorporating a grating structure that efficiently out-couples high-*k* states from the HMM structure.

In the ideal structure approximated by EMT hyperbolic dispersion corresponds to an infinite PDOS, however, in a realistic structure, having a finite number of layers of finite thickness and including losses from the metal, the PDOS is discretized into plasmonic modes. Fig. 1c shows the calculated local PDOS for an emitter placed 15*nm* below a multilayer structure composed of 8 alternating layers of Ag and $Al_2O_3$, with thickness of 15*nm* each as a function of normalized wave-vector parallel to the layers ($k_x/k_0$). The emitter is taken to have a dipole moment oriented normal to the multilayer film interfaces and is calculated using a dyadic Greens function approach [19]. The three bright bands around $k_x/k_0 = 5$, termed high-*k* modes or plasmonic Bloch modes are due to coupling of surface-plasmon-polaritons at each metal/dielectric interface. These modes have a strong electric field intensity that overlaps with emitters placed near the metamaterial thus resulting in an enhanced PDOS. This enhancement gives rise to preferential emission into high-*k* modes which channel the spontaneous emission along the resonance cone within the metamaterial [17].

Finite element method (FEM) based simulation of the electric field generated by a vertical electric dipole emitting at 635*nm* and placed 30*nm* below the HMM shows preferential emission into the structure in the shape of a resonance cone (Fig. 2a). The choice of emission wavelength for simulation was dictated by the quantum dots (QDs) used in our experiments, however, the structure supports broadband enhancement in spontaneous emission as was previously shown for similar composites [5]. In the cross-section of the HMM shown in Fig. 2a it is noticeable that most of the field is strongly confined inside the HMM and becomes evanescent at the top interface resulting in almost no detectable emission in the far field. In order to translate the evanescent field intro propagating waves a bulls-eye grating structure made of Germanium (Ge) was designed with a half-period Δ= 150*nm* and height of 17$nm$. Ge was selected due it's compatibility with the other materials used in the HMM fabrication and a high index of refraction which allows better matching with the HMM's high-$k$ modes, an advantage which out-weighs the material losses of Ge in the visible spectral range. The circular symmetry of the grating design was chosen specifically for directional out-coupling from randomly oriented dipoles. See Supporting Information for further details about the design of the grating.

The inset in Fig. 2a shows a cross-section of the near field emission from a vertical dipole below the 4P structure with a grating where two lobes extending into the far field can be easily distinguished. The angular resolved far field (Fig. 2b) shows a narrow emission cone with a half angle of ~$20^0$ into free space.

The optical quality of the layers is highly important in order to obtain bulk plasmons-polariton modes and avoid local scattering. A silver layer with thickness of less than 20$nm$ tends to form percolated films when deposited on dielectric materials. It has been shown that a seed layer of Ge bonds readily with Ag and results in an optically smooth surface with very low roughness [20,21]. We used a Ge seed layer (<2$nm$) which allows fabrication of ultra-smooth thin silver films without significantly altering the overall optical properties of the HMM. Atomic Force Microscope (AFM) image of a single 15$nm$ Ag layer deposited on top of the Ge seed layer deposited via electron beam evaporation is shown in Fig. 3a. The Ag film is found to have maximum peak to valley roughness of 0.5$nm$. After establishing the smoothness of the thin Ag layers, the active HMM structures were fabricated. The first layer consists of CdSe/ZnS core/shell QDs embedded in a poly-methyl-methacrylate (PMMA) matrix on top of pre-cleaned glass cover slip deposited via spin coating. Next, four period (4P) and one period (1P) structures (a single period consists an $Al_2O_3$ layer and a Ag layer with Ge seed) were grown on top of the QDs matrix and capped with a thin layer (5*nm*) $Al_2O_3$ in order to protect the last Ag layer from oxidation. The smoothness and continuity of the multiple layers composing the 4P structure has been confirmed by cross-sectional TEM as shown in Fig. 3b. As can be seen from the TEM image the final structure was less than 50% fill fraction. However calculations based on EMT and FEM showed that even with variations in the layers thicknesses, the structure was still strongly hyperbolic in the visible spectral range (see Supporting Information Section 2). Finally, a layer of 100*nm* PMMA resist was spin coated on top of the capping layer and grating structures with 150, 200, 265 and 300*nm* half-periods were etched in the resist by e-beam lithography (Fig. 3c). The PMMA grating is then used as a mask for depositing Ge and realizing the high index bulls-eye gratings.

Photoluminescence measurements were carried out using an inverted confocal microscope setup coupled to an avalanche photo diode detector with a 530*nm* low-pass filter and pumped with a solid state pulsed laser at a wavelength, λ=440*nm*. Fig. 4a shows the photoluminescence image of a Δ= 150*nm* grating on 4P showing strong emission from the grating with a bright center. Finite-difference-time-domain based simulations of emission from randomly oriented dipoles show that the out-coupling efficiency decreases dramatically for emitters located away from the center axis of the bulls-eye which suggests an explanation to the bright emission feature in the center of the bulls-eye observed experimentally. Figure 4(b) shows the spectrum normalized far-field for dipoles located 15*nm* below the multilayer structure, at various locations relative to the center bulls-eye axis. For emitters modeled as Hertzian dipoles there are three independent orientations for the dipole moment: (1) along the bulls-eye axis, (2) normal to the bulls-eye axis and normal to the tangent of the rings, and (3) normal to the bulls-eye axis and parallel to the tangent of the rings. In experimental conditions, the measured intensity is an ensemble average of these three orientations. Shown in Fig. 4c is the experimentally measured ratio between far field intensity at the center of the grating to that of the background for 4P and 1P structures along with the theoretical simulation. The $\Delta = 150 nm$ grating is shown to provide the best out-coupling for both the 4p and the 1P structure. While the single period supports only two surface-plasmon-polaritons with different charge parity, the 4P structure supports two surface states, and additional high $k$-states (3 in this case). The number of high-$k$ modes is dictated by the number of confined dielectric layers. The excellent agreement found between the experiments and the theoretically simulated light-out-coupling efficiency, substantiates that it is the existence of the high-$k$ modes which is responsible for the *tenfold* ratio between grating and background in the 4P HMM. In addition to the steady state results shown here, the HMM structure also showed the expected reduction in spontaneous emission lifetime due to the increase in PDOS similar to ones previously reported [5] (see Supporting Information Section 3 and figure 3)

In summary, we demonstrate efficient light extraction from an active HMM supporting high-$k$ modes embedded with quantum dots. A high refractive index Ge bulls-eye grating structure fabricated on top of the active HMM was used to enhance the light extraction from the high-$k$ modes. In addition to out-coupling the extracted light also shows directionality. Control of directionality and extraction of resonance cone emission from active HMM is an important step to achieving practical photonic devices such as sub-wavelength lasers, superluminescent LEDs and single photon guns.

**Materials & Methods**

CdSe/ZnS core/shell quantum dots purchased from NN Labs in Toluene were added to PMMA and Toluene solution in 25% by volume concentration and then spin-coated on pre-cleaned glass cover slips. $Al_2O_3$, Ge, and Ag layers were deposited by e-beam evaporation using a Lekser PVD 75 Thin Film System under pressure $< 1 \cdot 10^{-5}$ *Torr*. PMMA A2 was spin coated at 7000RPM and etched with JEOL JBX6300-FS e-beam lithography system. The pattern was developed in 3:1 MIBK and Isopropanol for 60 seconds. 17*nm* of Ge were e-beam evaporated on top. The PMMA was then removed by a lift-off procedure in Acetone solution. **Imaging.** Fluorescence microscopy and lifetime measurements were performed on an inverted confocal microscope setup coupled to a diode pumped solid state laser delivering 440*nm,* 90*p*sec pulses, at 8MHz repetition rate. The sample's luminescence was spectrally separated from the laser by a Semrock RazorEdge 532 low pass (LP) filter and detected by an MPD Picoquant Avalanche Photodiode coupled to a PicoHarp 300 time analyzer.

**Notes**

The authors declare no competing financial interest.

**ACKNOWLEDGMENTS**

This work was supported by the Army Research Office through Grant NO. W911NF1310001. Research carried out in part at the Center for Functional Nanomaterials, Brookhaven National Laboratory, which is supported by the U.S. Department of Energy, Office of Basic Energy Sciences, under Contract No. DE-AC02-98CH10886. The authors thank Mircea Cotlet and Huidong Zang for assistance with imaging and Yun Yu for help with AFM measurements.

**FIGURES**

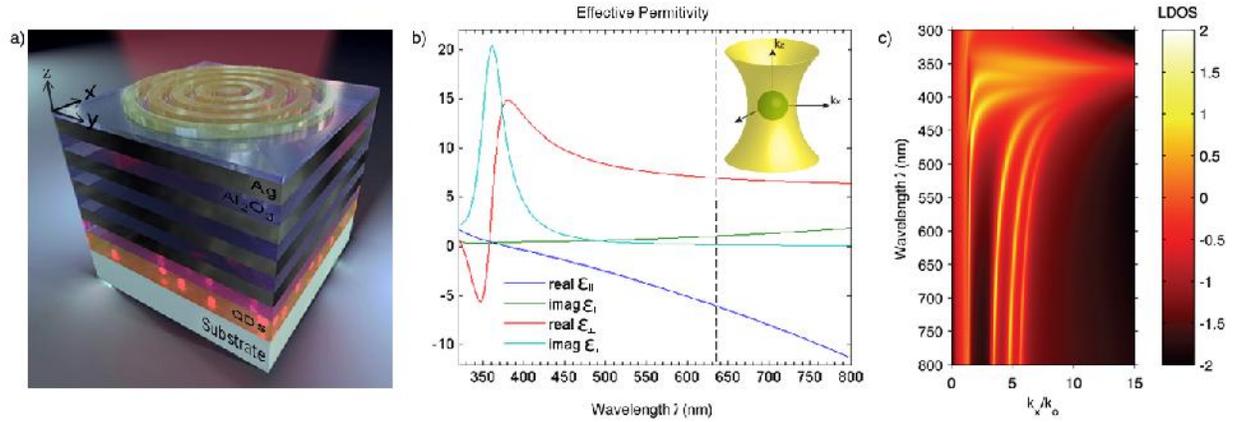

**Figure 1.** (a) Schematic of the active HMM with bulls-eye grating (b) Permittivity of the structure as calculated from Effective Medium Theory based on ellipsometrically determined dielectric constants of $Al_2O_3$ and thin film Ag and thicknesses determined from cross-sectional TEM. The dotted line corresponds to the emission maximum of the quantum dots used in the present work (635 *nm*). Inset: The hyperbolic iso-frequency surface of an ideal HMM overlaid on the spherical iso-frequency surface of free space. (C) Local Photonic Density of States for the HMM structure consisting of eight alternating 15 *nm* thick layers of Ag and $Al_2O_3$.

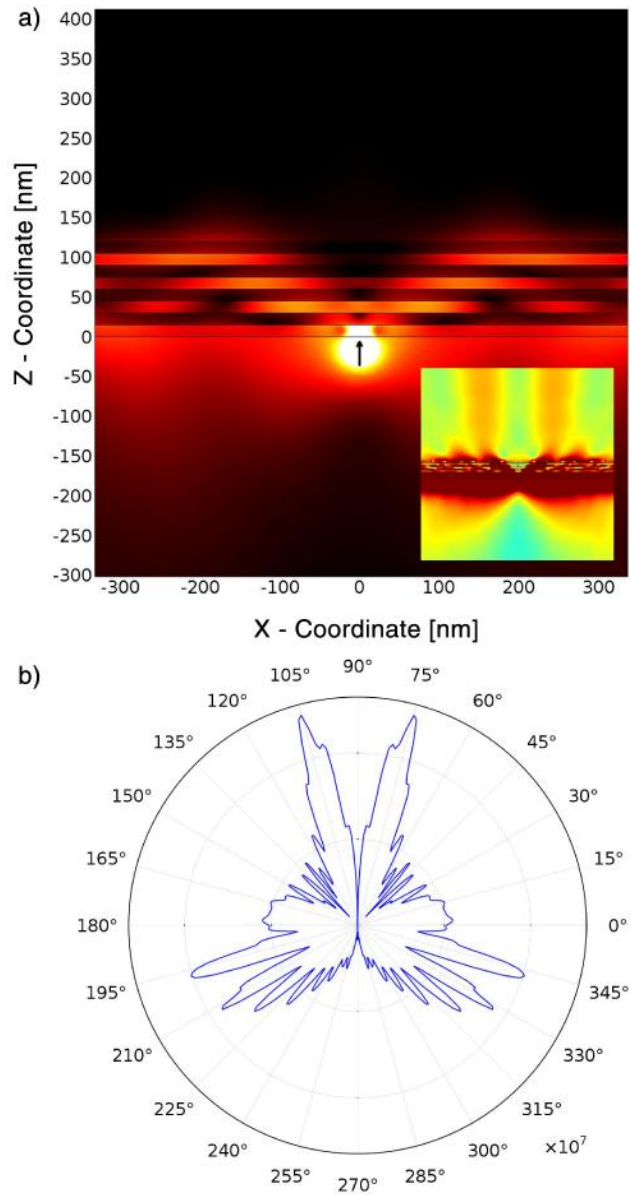

**Figure 2.** (a) Simulation of the electric field generated by a vertical dipole placed in the near field of the HMM. Black arrow represents the vertical dipole. Inset: near field in the presence of Ge bulls-eye grating resulting in out-coupling of the evanescent field. (b) Far field of the dipole emitter near the HMM with Ge bulls-eye grating.

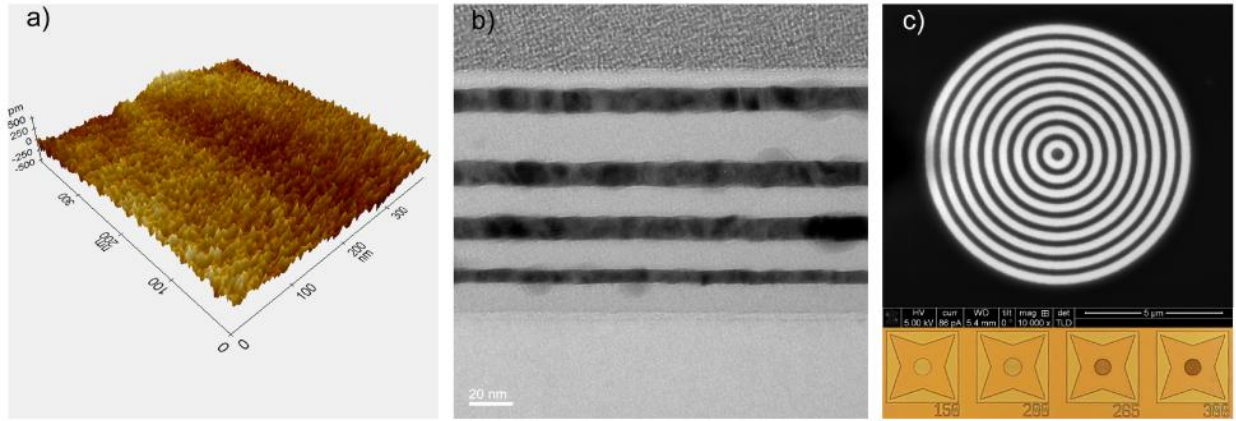

**Figure 3.** (a) Atomic Force Microscope (AFM) image of the surface of 15*nm* Ag over a 2*nm* Ge seed layer deposited via electron beam evaporation showing approximately 0.5*nm* peak to valley roughness. (b) Cross-sectional Transmission Electron Microscope (TEM) image of an identical 4P HMM structure grown on silicon showing smooth continuous films of both Ag and $Al_2O_3$. The top layer is palladium that was deposited to aid the cross-sectioning. (c) Scanning electron microscope (SEM) image of bulls-eye grating along with an optical microscope image showing the grating structures with different periodicities.

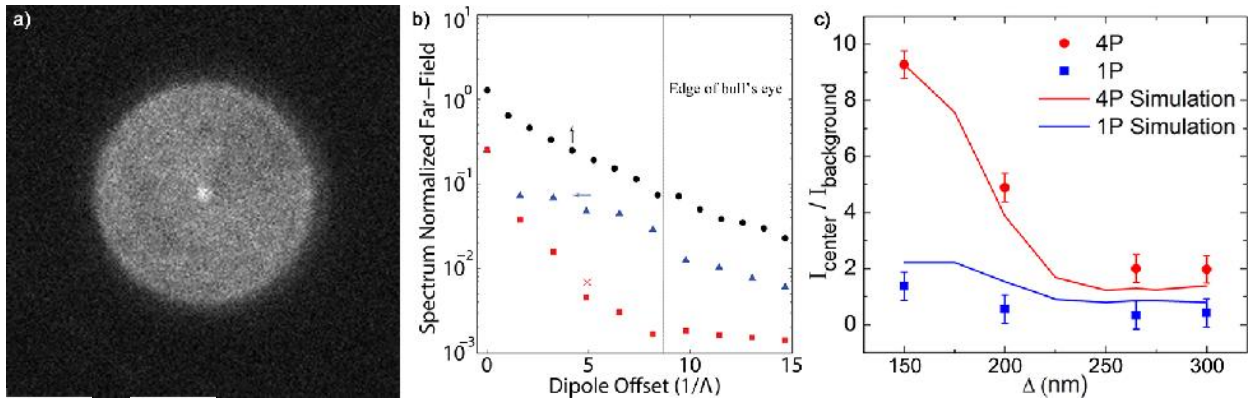

Figure 4. (a) Confocal microscope image of emission from the active HMM with Ge bulls-eye grating with half period, Δ=150nm. The measurements were carried out over a scan area of $20\mu m \times 20\mu m$. Details of the measurement can be found in the methods section. (b) FEM simulation of out-coupled power for a single dipole offset from grating center. vertical orientation (circles), orientation in direction of the offset (rectangles), orientation in the orthogonal in-plane direction (squares). (c) Experimentally measured ratio between the intensity at the center of the grating to that of the background (regions outside the grating) for the 4P and the 1P structures with different grating periods. Also shown in solid lines are the theoretically simulated ratios for the two cases using ensemble approximation for dipole located at an average distance of ~ 30 nm below the first $Al_2O_3$ layer.

## ASSOCIATED CONTENT

**Supporting Information**

Simulations for out-coupling as a function of grating half-period, Simulation of supported modes for measured layers thicknesses, and Liftime measurements.

## AUTHOR INFORMATION

**Corresponding Author**

E-mail: vmenon@qc.cuny.edu

# Supporting information

# Directional out-coupling from active hyperbolic metamaterials

*Tal Galfsky[1,2*], Harish N.S. Krishnamoorthy[1,2], Ward D. Newman[3], Evgenii Narimanov[4], Zubin Jacob[3], Vinod M. Menon[1*]*

[1]Department of Physics, Queens College, City University of New York (CUNY) and Center for Photonic and Multiscale Nanomaterials, Flushing, NY 11367, USA.

[2]Department of Physics, Graduate Center, CUNY, New York, NY 10016, USA.

[3]Department of Electrical and Computer Engineering, University of Alberta, Edmonton, AB T6G 2V4, Canada.

[4]Birck Nanotechnology Center, School of Computer and Electrical Engineering, Purdue University, West Lafayette, Indiana 47907, USA.

* Corresponding authors: vmenon@qc.cuny.edu

## 1. Grating design

Fig 1a. graphs the upwards portion of an emitter's power as a function of the grating half-period within a $30^0$ half angle from the normal to the HMM surface for a vertical dipole at a distance of 5nm below the HMM and emitting at 635nm. In the region between $\Delta = 150 - 175 nm$ a significant portion of the dipole power is directed into a very narrow emission cone. However, luminescent dipole sources like dyes or quantum dots (QDs) have a broad emission (~50$nm$ at FWHM) and are randomly oriented. Hence one needs to take into account the output of both the vertical and horizontal dipoles as well as integrate over the emitter's spectrum in order to simulate the realistic scenario, as seen in Fig 1b for $\Delta = 150 nm$.

## 2. Experimental Structure Details

Transmission Electron Microscopy has been used for high resolution measurements of the thickness of metal (M) and dielectric (D) layers in the active HMM (see Fig. 2a). The thicknesses starting from the bottom are: $D_1$ = 13$nm$, $M_1$ = 16$nm$, $D_2$ = 23$nm$, $M_2$ = 7.6$nm$, $D_3$ = 17.8$nm$, $M_3$ = 7.6$nm$, D4 = 15$nm$, $M_4$ = 7.3$nm$, $D_5$ (capping layer) = 6.5$nm$. Every metal layer is composed of Ag with a 1 to 2$nm$ Ge seed layer. Mode analysis simulations were preformed to determine the existence and location of high-$k$ modes (Fig. 2b). At a wavelength of 635$nm$ there are three high-$k$ modes. The first located at $k_x/k_0$ = 2.8, the second at $k_x/k_0$ = 5, and the highest third and therefore the most highly confined at $k_x/k_0$ = 7.24.

## 3. Lifetime measurements

Lifetime measurements displayed in Fig. 3 were conducted on an inverted confocal microscope setup coupled to a diode pumped solid state laser delivering $440 nm$ $90 psec$ pulses, at 8MHz repetition rate. The sample's luminescence was spectrally separated from the laser line by a Semrock RazorEdge 532LP filter and detected by an MPD Picoquant Avalanche Photodiode coupled to a PicoHarp 300 time analyzer. Spontaneous emission kinetics of CdSe/ZnS Core Shell Nanocrystals were measured for a single period (1P) and four periods (4P) of the dielectric-metal composite without the bottom PMMA and with quantum dots (QDs) directly spin-coated directly on top of the Al2O3 spacer layer (5-7 nm thickness). The reason for making a sample specifically for lifetime measurements is that in order to see a clear influence of the enhanced photonic density of states (PDOS) in a HMM the emitters should be confined as best as possible to a single uniform distance from the surface since the coupling efficiency to high-*k* modes changes as a function of distance to the structure[1]. In the sample containing QDs in PMMA the distance distribution of emitters in PMMA results in a wide-spread distribution of lifetimes while spin coating QDs on top of the spacer layer confines the emitters to a plane at ~15nm from the first Ag layer (accounting for the QDs physical size) and results in good coupling as expressed by the lifetimes (glass = 25*ns*, 1p = 11*ns*, 4p = 5*ns*). The lifetime ratios are in good agreement with previous reports [2,3].

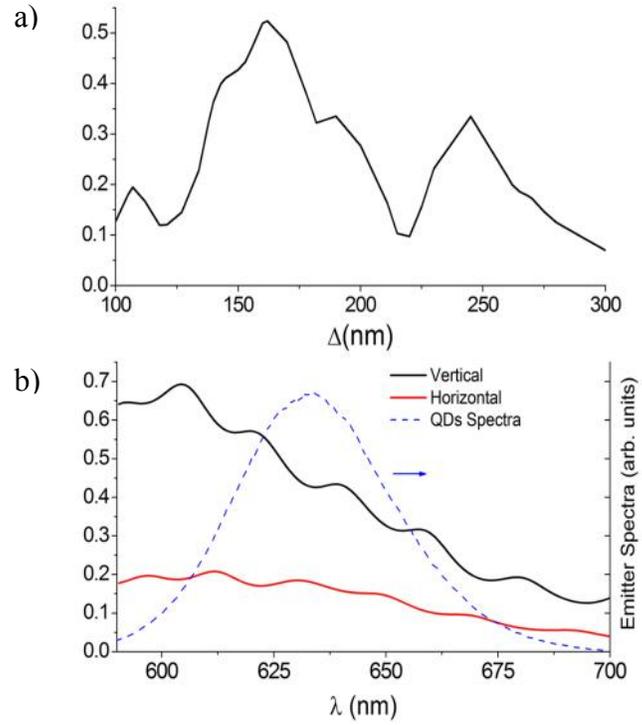

**Fig. 5.** Out-coupled power within a 30° degrees half-angle normalized to the total power emitted by a dipole in PMMA. (a) as a function of the grating half period, $\Delta$, for a vertical dipole. (b) for vertical (upper line) and horizontal (lower line) dipoles with $\Delta = 150nm$ as a function of wavelength. QDs spectra in dashed line.

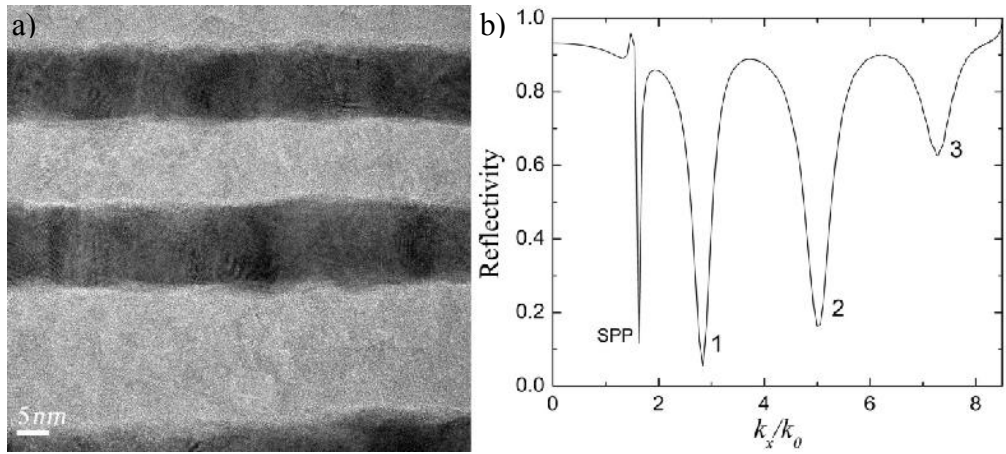

**Fig. 6.** (a) TEM image of layers in HMM. (b) Mode structure calculated with the measured layers thicknesses as determined from simulated reflectivity for wavelength $\lambda = 635nm$.

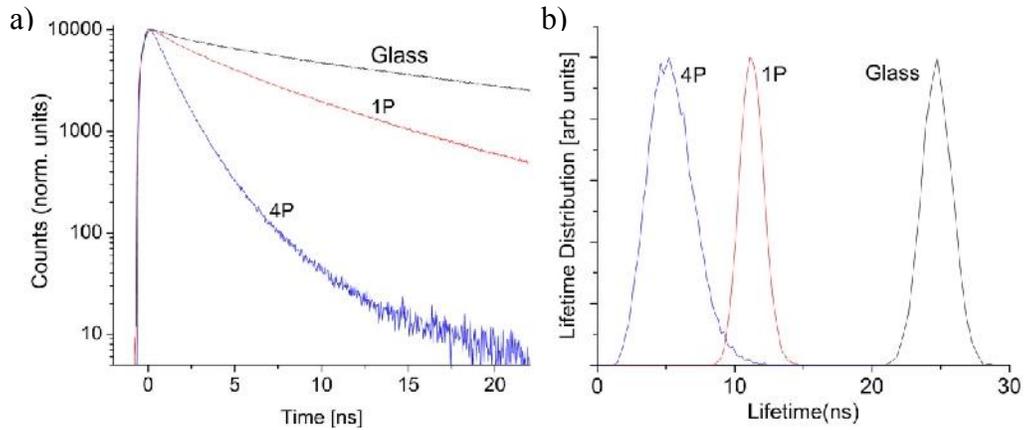

**Fig. 7.** Emission kinetics for (a) QDs on top of glass, 1P, and 4P structures (b) Lifetime distributions.